\documentstyle[epsfig,12pt,amstex]{article}
\begin{document}
\title{ QCD, conformal invariance and the two Pomerons}
\author{S. Munier and R. Peschanski \\ 
CEA, Service de Physique Theorique  CE-Saclay \\ F-91191
Gif-sur-Yvette Cedex, France \\}
\maketitle
\begin{abstract}

Using the solution of the BFKL equation including the leading and  subleading conformal spin components, we show how the conformal invariance underlying the leading $\log (1/x)$ expansion of perturbative QCD leads  to elastic amplitudes described by two effective Pomeron singularities.  One Pomeron is the well-known "hard"  BFKL leading singularity while the new one appears from a shift of the  higher conformal spin BFKL singularities from  subleading to leading position. This new effective singularity is compatible with the "soft" Pomeron and thus, together with the "hard" Pomeron,  meets  at large $Q^{2}$ the "double  Pomeron" solution  which has been  recently conjectured by Donnachie and Landshoff.
\end{abstract}

\section{Introduction : two Pomerons ?} 
In a recent paper  \cite{donnadrie} the
conjecture was made that not one, but two Pomerons could
coexist. This proposal is based on a  description of data for the proton
singlet structure function $F\left( x,Q^{2}\right) $ in a wide range of  $x\left(
<0.7\right) $ and all available $Q^{2}$ values (including also the charm structure function and  elastic photoproduction of $J/\Psi $ on the
proton). The singlet structure function reads\begin{equation}
F\left( x,Q^{2}\right) =\sum\limits_{i=0}^2F_{i}\left( x,Q^{2}\right) =\sum\limits_{i=0}^2f_{i}\left( Q^{2}\right) x^{-\epsilon_{i}}, 
\label{1}
\end{equation}
corresponding    \cite {donnadrie,parametrization}
to the sum of three contributions, namely a ``hard'' Pomeron contribution with a fitted intercept $\epsilon_{0}=.435, $ a ``soft'' Pomeron exchange, as seen in soft hadronic
cross-sections with a fixed   intercept $\epsilon_{1}=0.0808, $ and a secondary Reggeon singularity necessary to describe the larger $x$ region with intercept fixed at $\epsilon_{2}=-.4525.$ The ``hard'' Pomeron is in particular needed to describe  the strong rise
of $F$ at small $x$ observed at HERA  \cite{adloff}. The key observation
of Ref.[1] is that the agreement with data can be obtained by assuming an opposite
 $Q^{2}$-behaviour for the two Pomeron contributions in formula (1). Indeed, for $Q^{2}>10 \ GeV ^{2},$ $%
f_{0}\left( Q^{2}\right) $ is increasing and $f_{1}\left( Q^{2}\right) $
decreasing (the precise
parametrizations \cite{parametrization} are given in a Regge theory framework).

This picture is suggestive of a situation where the ``soft'' and 
 ``hard'' Pomerons are not one and the same object but two separate Regge singularities with rather different intercept and $Q^2$ behaviour. The ``hard'' Pomeron may be  expected to be  governed by perturbative QCD evolution equations.\ Indeed, at
small $x,$ a Regge singularity is expected to occur as a solution of the
BFKL equation \cite{bfkl} corresponding to the resummation of the leading $%
\left( \bar{\alpha}\ln 1/x\right) ^{n}$ terms in the QCD perturbative
expansion, where $\bar{\alpha}= \frac {\alpha N_c} {\pi}$ is the (small) value of the coupling
constant of QCD. The intercept
value  is predicted to be 
$\epsilon_{0}= 4\bar{\alpha}\ln 2.$ 
It is interesting to note that the phenomenological fit for the hard Pomeron
in Ref.\cite{donnadrie} corresponds to a reasonable value for $\bar{\alpha}%
\left( \approx .15\right) .$ The goal of the present paper is
to show that  the global conformal
invariance of the BFKL equation \cite{lipatov} leads to a natural mechanism
generating both the ``hard'' and ``soft'' Pomeron singularities.

The plan of the paper is as follows: in Section {\bf 2},  using the BFKL equation and the set of its conformal-invariant components, we exhibit  the
phenomenon generating sliding singularities. In
Section {\bf 3}, we explicitly describe the two-Pomeron
configuration obtained from the ``sliding'' mechanism. In section {\bf 4} we confront the resulting effective singularities
with the parametrization of  \cite{donnadrie} and
discuss some expectations from non-perturbative corrections at small $Q^2.$ Finally, in section {\bf 5}, we discuss some phenomenological and theoretical  implications of our QCD two-Pomeron mechanism.
 
\section{The ``sliding'' phenomenon}
\quad Let us start with the  solution of the BFKL
equation expressed in terms of an expansion over the whole set of conformal spin components  \cite{lipatov}. For structure functions, one may write (using the notation $Y=\ln 1/x$): 
\begin{equation}
F\left( Y,Q^{2}\right) =\sum\limits_{p=0}^{\infty }F_{p}\left(
Y,Q^{2}\right) =\sum\limits_{p=0}^{\infty }\int_{1/2-{\rm i}\infty }^{1/2+%
{\rm i}\infty }d\gamma \left( \frac{Q}{Q_{0}}\right) ^{2\gamma }e^{
\bar{\alpha}\chi _{p}\left( \gamma \right) Y}f_{p}\left( \gamma
\right) ,  \label{2}
\end{equation}
with 
\begin{equation}
\chi _{p}\left( \gamma \right) =2\Psi \left( 1\right) -\Psi \left(
p+1-\gamma \right) -\Psi \left( p+\gamma \right)  \label{3}
\end{equation}
and
$Q_{0},$ being some scale characteristic of the target (onium, proton,
etc...). $\chi _{p}(\gamma)$
is the BFKL kernel eigenvalue corresponding to the $SL(2,{\cal C})$
unitary representation  \cite {lipatov} labelled by the  conformal spin
$p.$ It is to be noted that the $p=0$ component
corresponds to the dominant ``hard'' BFKL Pomeron. Usually the $%
p\neq 0$ components, required by conformal invariance\footnote {In the following, we will stick to integer values of $p$  since half-integer spin components exist but do note contribute to elastic cross-sections \cite{navelet}.} but subleading by powers of the energy, are omitted with respect  to the leading logs QCD resummation. They are commonly
 neglected in the phenomenological discussions. We shall see that they may play an important r\^ ole, however.

The couplings of the BFKL components to external sources are taken into account by 
the weights $f_{p}\left( \gamma \right) $ in formula (2). Little is known
about these functions and we shall treat them as much as possible in a model independent way. For instance, they should obey some general
constraints, such as a behaviour when $\gamma \rightarrow \infty $ ensuring
the convergence of the integral in (2). We will see that some extra analyticity constraints will appear in the context of the two Pomeron problem\footnote{%
Note that a general constraint on the coupling of the BFKL kernel to
external particles is coming from gauge invariance \cite{lipatov}. We checked
that this constraint is rather weak in our case, and not relevant to the discussion.}.

The key observation leading to the sliding phenomenon starts by  considering the successive derivatives of the kernels $\chi _{p}\left(
\gamma \right) .$ One considers the following suitable form: 
\begin{eqnarray}
\chi _{p}\left( \gamma \right) &\equiv &\sum\limits_{\kappa =0}^{\infty }%
\left\{\frac{1}{p+\gamma +\kappa } +\frac{1}{p+1-\gamma +\kappa }-\frac{2}{\kappa +1}\right\}
\nonumber \\
&&  \nonumber \\
\chi _{p}^{\prime }\left( \gamma \right) &\equiv &-\mathop{\displaystyle\sum}\limits_{\kappa }
\left\{\frac{1}{\left( p+\gamma +\kappa \right) ^{2}}-\frac{1}{%
\left( p+1-\gamma +\kappa \right) ^{2}}\right\}  \nonumber \\
&&  \nonumber \\
\chi _{p}^{\prime \prime }\left( \gamma \right) &\equiv &2\mathop{\displaystyle\sum}\limits_{\kappa }
\left\{\frac{1}{\left( p+\gamma +\kappa \right) ^3}+\frac{1}{%
\left( p+1-\gamma +\kappa \right) ^{3}}\right\}\ .
\label{4}
\end{eqnarray}
As obvious from (4), the symmetry $\gamma \Longleftrightarrow 1\!-\!\gamma $
leads to a maximum at $\gamma =1/2$ for all $p,$ and thus to a saddle-point
 of expression (2) at $\Re e(\gamma) =1/2$ for ultra asymptotic values of $Y.$ The saddle-point approximation  gives
\begin{equation}
F\left( x,Q^{2}\right) \vert_{Y\rightarrow \infty }
\approx 
\left( \frac{Q}{Q_{0}}\right)\mathop{\displaystyle\sum}\limits_{p=0}^{\infty }\frac {f_{p}\left( \frac 12\right)}
{\sqrt {\pi \bar{\alpha} \chi _{p}^{\prime \prime }\left( \frac12\right) Y}}\ e^{\bar \alpha \chi _{p}\left( \frac12\right) Y}. 
\label{5}
\end{equation}
The $Q$-dependent factor corresponds to a common   anomalous dimension $\frac 12$ for all $p.$ Note that the known $Q$-dependent ``$k_T$-diffusion'' factor is absent in this ultra-asymptotic limit.

The  series of functions of $Y$ is such that
only the first term has intercept $ \bar \alpha\chi _{p}\left( \frac 12 \right) $ larger than $0.$ Indeed, 
\begin{align}
\chi _{0}\left(\frac 12\right) &=4\ln 2 \approx 2.77 \nonumber\\
\chi _{1}\left(\frac 12\right) &=\chi _{0}\left( \frac12\right)
-4\approx -1.23 \nonumber \\ 
\chi _{p+1}\left( \frac12\right) &<\chi _{p}\left( \frac12\right)
<...<0 ,\ \  p\geq 1. 
\label{6}
\end{align}
This ultra asymptotic result is  the reason why the conformal spin components with $p>0$
are generally neglected or   implicitly taken into account by   ordinary secondary Regge singularities  with intercept less than $0.$
However, at large enough values of $Q^2$ and even for very large $Y,$ a
sliding phenomenon moves away the singularities corresponding to these conformal
spin components, leading to a very different behaviour from  (5). Indeed, the sliding mechanism is already known \cite {bartelo,npr} to generate the diffusion factor of the leading $p=0$ component. However it has an even more important effect on the higher spin components as we shall discuss now.

The sliding mechanism is based on the fact that $\chi _{p}^{\prime \prime}\left(\frac 12\right),$ the second derivative of the kernels at the asymptotic saddle-point value,  becomes in absolute value very small when $p \geq 1,$ in such a way that the real saddle-points governing the
integrals of formula (2) are considerably displaced from $\gamma =1/2.$
Indeed, considering the expansions (4), one has:
\begin{align}
\chi _{0}^{\prime \prime }\left( \frac 12\right) &=28\zeta (3)\approx   33.6\nonumber
\\
\chi _{1}^{\prime \prime }\left( \frac 12\right) &=28\zeta \left( 3\right)
 -32\approx  1.66  \nonumber \\
 1.66>...&>\chi _{p}^{\prime \prime }\left(
\frac 12\right) >\chi _{p+1}^{\prime \prime }\left(
\frac 12\right) >0 ,\  p \geq 2\ .
\label{7}
\end{align}
For the $p=0$ component, the corresponding integral in (\ref {2}) can be evaluated by a saddle-point in the vicinity of $\gamma = \frac 12,$ and gives the diffusion factor $\exp\left(-\log^2(Q/Q_0)^2/2\bar \alpha Y \chi _{0}^{\prime \prime }\left( \frac 12\right)\right) .$ Considering the rapid decrease by a factor 20 of the modulus of the second derivative for $p=1,$ it is easy to realize that, for components $p \geq 1 ,$ it is no more justified to evaluate the integrals in the vicinity
of $\gamma = \frac 12,$ the real saddle-point being  away from this value. We shall make the correct evaluation in the next section.

\section{The ``sliding'' mechanism}
Let us consider the $F_{p}$ component of the summation (2) in the following way: For 
each value of $\left( Y,\ln \frac{Q^{2}}{Q_{0}^{2}}\right) ,$ we  compute  the {\it effective intercept}  (in units of $\bar\alpha$) ${\displaystyle{\partial \ln F_{p} \over \bar \alpha \partial Y}}$  displayed  as a function of the {\it effective anomalous dimension} ${\displaystyle{\partial \ln F_{p} \over \partial \ln Q^{2}}}=\gamma_{c}.$  Our observation is that, for any weight $f_{p} \left( \gamma \right)
 $ in formula (2), the resulting set of points accumulates near the curve $\chi _{p}\left( \gamma \right).$ This result is valid provided a saddle-point  dominates the
integral. 

The proof comes as follows: If a saddle point $\gamma _{c}$ dominates the
integral (2) for $F_{p}\left( Y,Q^{2}\right) ,$  the saddle-point
equation 
\begin{equation}
 \frac{\partial \ln F_{p}}{\partial
\gamma _{c}}= 2\ln \left( Q/Q_0\right) ^{2}+\bar{\alpha}Y\chi _{p}^{\prime }\left( \gamma
_{c}\right) +\left[ \ln f_{p}\left( \gamma _{c}\right) \right] ^{\prime }=0 \label{8}
\end{equation}
is verified and  the resulting integral is approximated by
\begin{equation}
F_{p}\left( Y,Q^{2}\right) \approx \frac {
 \left( Q/Q_{0}\right) ^{2\gamma _{c}}e^{\bar{\alpha}\chi
_{p}\left( \gamma _{c}\right) Y}\ f_{p}\left( \gamma _{c}\right)}
{
\left\{2\pi \left( \bar{\alpha}Y \chi
_{p}^{\prime \prime }\left( \gamma _{c}\right) +\left[ \ln f_{p}\left(
\gamma _{c}\right) \right] ^{\prime \prime }\right)\right\}^{\frac 12}}\ .  
\label{9}
\end{equation}
Neglecting in (9) derivatives of the slowly
varying saddle-point prefactor $\left\{...\right\}^{-\frac 12},$ one may write 
\begin{eqnarray}
\frac{d\ln F_{p}}{\bar{\alpha}dY} &=&\frac{\partial \ln F_{p}}{%
\partial \gamma _{c}}\times \frac{d\gamma _{c}}{\bar{\alpha}dY}+\frac{%
\partial \ln F_{p}}{\bar{\alpha}\partial Y}=\frac{\partial \ln F_{p}}{\bar{%
\alpha}\partial Y }\equiv \chi _{p}\left( \gamma _{c}\right)  \nonumber
\\
&&  \nonumber \\
\frac{d\ln F_{p}}{d\ln Q^{2}} &=&\frac{\partial \ln F_{p}}{\partial
\gamma _{c}}\times \frac{d\gamma _{c}}{d \ln Q^{2}}+\frac{\partial \ln F_{p}}{%
\partial \ln Q^{2}}=\frac{\partial \ln F_{p}}{\partial \ln Q^{2}}\equiv
\gamma _{c},
\label{10}
\end{eqnarray}
where one uses the saddle-point equation (8) to  eliminate the
contributions  due to the implicit dependence  $\gamma _{c}\left(
Y,Q^{2}\right) .$ This proves our statement.

Interestingly enough, the property (10) is valid for any weight $f_{p}\left( \gamma \right),$ and thus can be used to characterize the generic behaviour of the expression (2). The only condition is  the validity of a saddle-point approximation which is realized whenever $Q^2$ or $Y$ is large enough.

Let us discuss some  relevant example.  In Figs.1,2 we have plotted the result  of the numerical
integration  in expression (2) for $p=0,1,2,$ choosing $f_{p}\left( \gamma \right) \equiv \frac{1}{\cos \frac{\pi\gamma }{4}}.$ This weight is chosen in such a way that  the convergence properties of the integrands are ensured  and no extra singularity is generated for $\vert\gamma\vert < 2.$ Other weights with the same properties were checked to give the same results.
For comparison we also display the functions $\chi _{0}\left( \gamma \right) ,\chi_{1}\left( \gamma \right) $ and $%
\chi _{2}\left( \gamma \right) .$ Note that we have also included for the discussion the auxiliary branches of $\chi _{0}\left( \gamma \right)$ for the intervals
 $-1<\gamma<0$ and  $-2<\gamma<-1.$

The results both for $p=0$ (white circles) and $p=1,2$ (black circles) are displayed in Fig.1 for a fixed  large value of total rapidity $Y=10$ and various values of $\ln {\displaystyle{Q^{2} / Q_{0}^{2}}},$ while in Fig.2 they are shown for a fixed value of $\ln {\displaystyle{Q^{2} / Q_{0}^{2}}}=4$ and various  $Y.$  Indeed,
it is  seen on these plots that the saddle-point property (10) is verified, even for the auxiliary branches\footnote{In the case of the two auxiliary branches
considered in Figs.1,2, we have considered an integration contour shifted by one and two units to the left in order to separate  the appropriate contributions from the leading ones.}. The observed small  systematic shift  of the numerical results w.r.t.  the
theoretical curves $\chi (\gamma)$ is well under control. It is   related to the saddle-point prefactor in formula (9). 

By various verifications, we checked
that the results shown in Figs.1,2 are generic  if the following three
conditions are realized:

\bigskip
i) $Y$ or $\ln Q^{2}/Q_{0}^{2}$ are to be large enough $\left( \geq 2,3\right) $ to allow for a saddle-point method.

ii) $f_{p}\left( \gamma \right) $ is constrained to ensure the convergence and positivity of
the integrals of expression (2) in the complex plane.

iii) $f_{p}\left( \gamma \right) $ has no singularity for $\Re e(\gamma)
>-p.$ 
\bigskip

The striking feature of the results displayed  in Figs.1,2 is that, while remaining in vicinity of  the
curve $\chi _{p}\left( \gamma _{c}\right) ,$ $ {\displaystyle{d\ln F_{p} \over \bar{\alpha}dY}}$ and ${\displaystyle{d\ln F_{p} \over d\ln Q^{2}}}$ are shifted  away from the ultra asymptotic saddle-point point at $\gamma =1/2.$ Moreover, the  shift is larger if  the conformal spin $p$ is  higher.

Let us make a particular comment on the analyticity constraint iii). Obviously, the presence of a singularity at ${\cal R}e\gamma
>-p$ would prevent the existence of a shift. Indeed,
in Fig.3, we show the result for $f_{p}\left( \gamma \right) ={\displaystyle{\left( \gamma \cos \pi \gamma /4\right)^{-1}}}$ where we have explicitely violated the constraint iii) by a pole at $%
\gamma =0.$ As a result, the components  $F_{1}$ and $F_{2},$ remain still very close to their reference curves   $\chi_{1}\left( \gamma \right) $ and $\chi _{2}\left( \gamma \right) ,$ but they appear  ``sticked'' at the singularity point $\gamma =0.$ Thus  the relation (10) remains verified, but the sliding
mechanism is ``frozen'' by the singularity, as expected from  analyticity properties.

The main consequence of the sliding mechanism is to substantially modify the evaluation
of the sum (2) with respect to the ultra asymptotic expectation (5).
Indeed\footnote {Using various examples we found this result to be generic provided constraints i) -- iii) are verified.} the situation seen on Figs.1,2 is general:\ the first
contribution $F_{0}$ is subject to a rather small shift from $\gamma =1/2,$
while the $p=1$ component $F_{1}$ remains  at values where ${\displaystyle{d\ln F_{1} \over \bar{\alpha}dY}}$ is slighly above 1 and ${\displaystyle{d\ln F_{1} \over d\ln Q^{2}}}$ is below $-1/2.$ The 
higher components $F_{2}$ and a fortiori $F_{p>2}$ lie in
regions with  negative  effective intercept and lower and lower values of the effective
anomalous dimension. 

It is instructive to compare the results of Figs.1,2 for the $p=1$ component with those obtained for the auxiliary branches of the $p=0$ one. Though being situated in the same range of effective anomalous dimension $\gamma$ as the 
$p=1$ component, the first auxiliary branch gives  sensibly lower (and almost all negative) values of the effective intercept in the considered kinematical range. Thus, the corresponding contributions to the $p=0$ amplitude are subdominant in energy with respect to the spin 1 amplitude. The same property holds for the second auxiliary branch which stays subdominant with respect to 
the $p=2$ component which, in any case is itself subdominant with respect to $p=1$. 

Thus,  the mechanism we suggest for the two-Pomeron  scenario is the following: the r\^ole of the ``hard'' Pomeron is played (as it should be) by the component $F_{0},$ while the r\^ole of the ``soft'' Pomeron is played by the other components, principally the component with unit conformal spin $F_{1}.$   Here this mechanism is realized in a range $\left( Y,\ln Q^{2}/Q_{0}^{2}\right)$ where perturbative QCD (with resummation) is valid. Extrapolation to 
the non-perturbative domain will be discussed in the next section.

\section{Physical expectations}
\quad It is now worth discussing  our results, obtained from QCD and  conformal symmetry,
in the context of the
phenomenological analysis of paper  \cite{donnadrie}.  Our goal is not to identify the two approaches since the  theoretical conformal spin expansion (2) is only  valid in the perturbative QCD region at large $Y$ and $Q^{2},$ while the approach of paper  \cite{donnadrie} takes into account data 
in the whole range of  $Q^{2}.$ Nevertheless it is interesting to confront our resulting effective parameters with those obtained from the description of    paper  \cite{donnadrie}.

In Fig.4 we show a  plot comparing our results with those obtained from the two
Pomeron  components of  \cite{donnadrie} in  terms of the effective parameters as previously. In the case of the parametrization of paper   \cite{donnadrie}, the effective intercept and anomalous dimension  are easily identified as, respectively,  $ \epsilon _{i}$ and $d\ln f_{i}\left( Q^{2}\right) / d\ln Q^{2},$ see expression (1). In order to make contact with phenomenology, we have fixed $\bar \alpha = .15,$ and $Q_0 = 135  \ \mbox{MeV}.$ This last value is somewhat arbitrary but corresponds to  rather high values of $\ln \left( Q/Q_{0}\right)
^{2}$ in the physical range, justifying the existence of a significant saddle-point. In practice, in Fig.4, we have considered $Y=10$ and $\ln \left( Q/Q_{0}\right)^2=\left(4,6,8,10\right).$ The crosses in Fig.4 correspond to the effective parameters extracted from the parametrization  \cite{donnadrie} and the black dots to our numerical results of the integrals (2) for the same values of the kinematical variables. We performed the calculation with $f_{p}\left( \gamma \right) \propto 1/\cos \frac{\pi \gamma }{4},$ but checked the validity of the results for other weights (with similar analyticity properties, cf. section 3.)

The main thing to be noticed is the reasonable agreement between both results for large values of $Q^{2}$ corresponding to the direction of the arrows on the figure.  A few remarks are in order:

i) The leading ``hard Pomeron'' singularity obtained by our results is of the type  used e.g. in the phenomenological description of proton structure functions in the dipole model of BFKL dynamics  \cite {npr}. However the value of the coupling constant, chosen here to match with the determination of the hard component by  \cite{donnadrie}, is larger than in one-Pomeron fits  \cite {npr} and in better agreement with the original BFKL framework.

ii) The nonleading singularity is obtained in the correct range fixed by   \cite{donnadrie} to be given by the ``soft'' Pomeron  \cite {donland}. It is to be remarked that, while the ``hard'' Pomeron singularity is mainly fixed by the choice of $\bar \alpha,$ the nonleading one is a  result of the sliding mechanism. We thus find this feature to be model independent and related to the asymptotic conformal invariance of the input amplitudes.

iii) As also seen in the figure, the agreement is not quantitative, especially at lower $Q^2,$ since the results obtained from our formula (2) appear as {\it moving} effective singularities while those from paper  \cite{donnadrie} are, by definition, {\it fixed} Regge singularities.

Let us comment further on this important difference. In perturbative QCD, submitted to obey a renormalization group property, one expects in rather general conditions a scale-dependent evolution, different from the Regge-type of singularities, at least for the singlet channel \cite {rujula} \footnote {Note, however, the different perturbative approach of  \cite {indurain}.}. It is thus not surprising that the various components obtained from our approach show this characteristic feature, see Figs.1-4. On contrary, pure Regge singularities will correspond to fixed intercepts as shown in Fig.4 by the horizontal lines. 

We feel that  moving effective singularities will remain  a typical feature of the ``hard'' singularity at high $Q^2$, at least if perturbative QCD is relevant in this case. The situation is obviously different for  the ``soft'' singularity which   intercept is  fixed at the known ``universal'' value for soft interactions  \cite{donland}. The behaviour of the ``soft'' singularity when $Q^{2}$ becomes small is not determined in our perturbative approach. It only predicts that it will become dominant when $Q^2$ will approach and decrease below $Q_0^2,$ as indicated by the effective anomalous dimension. Non-perturbative QCD effects could thus be expected to stabilize the perturbative soft singularity at the known location of the phenomenological soft Pomeron\footnote {Another possibility \cite{bialas} could be a pole in the weight $f_p{(\gamma)}$ at a suitable position, but it would not be easily justified by a physical property like e.g. conformal invariance.}. Moreover, one would have to consider also the other higher conformal spin components.

Some  qualitative arguments can be added in favour of specific non perturbative effects for  conformal spin components. Indeed, the same reason leading to the sliding mechanism, namely the
smallness of $\chi _{p}^{\prime \prime }\left( \gamma \right) $ in the
vicinity of $\gamma =1/2,$ implies a large ``$k_{T}$-diffusion'' phenomenon 
 \cite{bartelo}. One typically expects a  range of ``$k_{T}$-diffusion''
for the gluon  virtuality scales building the spin component $F_{p}$ depending on $p$ as  $\left(\chi _{p}^{\prime \prime}\left(\frac 12\right)\right)^{-1}.$   Thus, while the contamination of non-perturbative unitarization effects
could be  limited for $F_{0}, $ it is expected to be strong for $%
F_{1}$ and the higher spin components $F_{p>1}.$  All in all, it is a consistent picture that the softer components obtained in a perturbative QCD framework at high  $Q^2$ are precisely those for which stronger ``$k_{T}$-diffusion'' corrections are expected. To go further would require a study of the low-$Q^{2}$ region, in particular of higher-twist contributions, which are outside the scope of our present paper \footnote {The known studies on higher-twists effects at low $x$  \cite {bartels1}  seems to show a different behaviour from the one obtained from the sliding mechanism of higher conformal spin components. This feature certainly deserves further study.}.

Concerning the physical meaning of the analyticity constraints imposed on the integrand factors $%
f_{p}\left( \gamma \right) ,$ they amount to discuss the conformal
coupling of the BFKL components to, say, the virtual photon and the proton (or, more generally other projectiles/targets). Leaving for future work the complete derivation of the conformal couplings to different conformal spins \cite {navelet2,appear}, let us assume that the coupling is spin independent. Interestingly enough an eikonal coupling to a  $q\bar{q}$ pair  \cite {muellertang} then appears to be forbidden, since it has  a pole at $\gamma =0,$ corresponding to the presence of the gluon coupling in the impact factor  \cite {mp}. However, considering the direct coupling through the probability distribution of  a virtual photon in terms of $q\bar{q}$ pair configurations  \cite{nikolaev}, we remark, following the derivation of \cite {mp}, that the pole due to the gluon coupling is cancelled with no other singularity at $\gamma=0.$ We explicitely
checked that we obtain very similar results to
those displayed in Figs.1--3 within this framework. Note that such a model ensures the positivity of the conformal spin contributions.

In our derivation, which follows from the conformal invariance of the BFKL equation, we have sticked to the case of a fixed coupling constant. It has been proposed  \cite {lipatov,braun,zoller} that the solution of  the BFKL equation, once modified in order to take into account a running coupling constant, leads to two, or more probably, a series of Regge poles instead of the $j$-plane cut obtained originally at
fixed $\bar{\alpha}.$ However, this solution with more than one Pomeron singularity does not ensure the specific $Q^2$ behaviour required by the analysis of \cite {donnadrie} and obtained by  the sliding mechanism. The running of the coupling constant, and more generally the results of the  next-leading BFKL corrections \cite{fadin},  modify the singularity structure could preserve the sliding mechanism.  Further study is needed in this respect.

\section{Conclusion and outlook}
\quad To summarize our results, using the full content of  solutions of the BFKL equation in a  perturbative QCD framework, and in particular their  conformal invariance, we have looked for the physical  consequences of the higher
conformal spin components of the conformal expansion on the problem of the Pomeron  singularites. We
have found, under rather general conditions, that the obtained pattern
of effective singularities leads to  two Pomeron  contributions,
one  ``hard'', corresponding to the ordinary conformal spin 0 component  and one ``soft'', corresponding to higher spin contributions, mainly spin 1. This situation meets, at least in the
large $Q^{2}$ domain, the empirical observation of Ref.\cite{donnadrie}
leading to a ``hard'' Pomeron  with leading-twist behaviour and a ``soft''
Pomeron  with higher-twist behaviour. It is interesting to note that the 
higher-twist behaviour we obtain corresponding to  the $p=1$ component is of higher effective intercept than the one which may be associated with the 
auxiliary branches of the ``hard'' component $p=0.$ Thus, there is no doubt that
the $p=1$ component behaviour is emerging from the other secondary BFKL contributions. However, its order of magnitude remains to be discussed \cite {appear}.

It is important to note that  the higher spin components rely on  the existence of an asymptotic global conformal invariance. This invariance has been proved to exist in the leading-log approximation. In the next-leading log BFKL calculations, It has been  recently advocated \cite {brod} to be preserved, at least  approximately. If this result is confirmed, and if the characteristics of the kernels are similar, the r\^ole  of the modified  higher conformal spin components is expected to be the same. Further tests of our conjecture also imply a study of the  specific couplings of the higher spin components to the initial states and an extension of the predictions to the non-forward diffractive scattering. Indeed, it has been recently shown \cite {levy} that the photoproduction of $J/\Psi$ gives evidence for no shrinkage of the Pomeron trajectory. Thus the two-Pomeron conjecture could also be borne out by considering non-forward processes.

 If confirmed in the future, the two-Pomeron  conjecture leads to further
interesting  questions, for instance:

- Can we built an Operator Product Expansion for the structure
functions, and thus higher-twist contributions, incorporating  the conformal invariance structure?

- Can we get some theoretical information on the physical ``soft'' Pomeron  by
considering  high-$Q^{2}$  indications given by perturbative QCD indications?

- Can we see some remnants of the specific conformal spin structure associated with the two Pomerons?

- The sliding mechanism appears as a kind of a spontaneous violation of asymptotic conformal invariance: can we put this analogy in a more formal way?

One interesting conclusion to be drawn from our study is that the matching of hard and soft singularities  could be very different from expectation. Usually, it is expected that  a smooth evolution is obtained from the hard to the soft region thanks to the increase of the unitarity
corrections to some ``bare'' Pomeron. By contrast, in the empirical approach of  \cite{donnadrie} and
in the theoretical sliding mechanism discussed in the  present paper, the
``hard'' and ``soft'' regions are essentially dominated by distinct
singularities, with only small overlap.  Clearly, this alternative deserves further phenomenological and theoretical studies. In particular, it has been suggested  \cite {bialas} to extend the study to (virtual) photon-photon reactions where the perturbative singularities and their specific coupling are
expected to be theoretically well-defined. For instance, if the eikonal coupling is confirmed as a characteristic feature of the (virtual) photon
coupling to the BFKL kernel, the sliding mechanism should not work for the spin 1 component and thus the would-be ``soft'' Pomeron   is expected to be absent from these reactions. Another case study is the Pomeron in hard diffractive reactions where the sliding mechanism, if present, could be different than for  total structure functions, and thus leading to a different balance of hard and soft singularities.

\bigskip  

{\bf ACKNOWLEDGEMENTS}

We want to thank the participants of the Zeuthen Workshop  on DIS at small $x,$ (``Royon meeting'', june 1998)  for fruitful discussions, among
whom Jeff Forshaw, Douglas Ross for stimulating remarks and particularly Peter Landshoff for provoking us with his and Donnachie's  conjecture. We are also  indebted to Andrzej Bialas and Henri Navelet for interesting  suggestions and comments.

\clearpage

{\bf FIGURE CAPTIONS}
\bigskip\bigskip

Fig.1. {\it Plot of effective intercept vs. effective dimension at fixed $Y$}

The effective intercept  $\partial \ln F_{p} / \bar \alpha \partial Y$ plotted  vs. the effective anomalous dimension $\partial \ln F_{p} / \partial \ln Q^{2}$ is  compared to the  $\chi _{p}\left( \gamma \right) $ functions for the 3 first conformal spin components ($p= 0,1,2$.). They are computed for 
 a fixed value of $Y=10$ and 4 values of $\ln {\displaystyle{Q^{2} / Q_{0}^{2}}}=\left\{ 4,6,8,10\right\} .$  The chosen weight in the integrals (2), see text,  is $f_{p}\left( \gamma \right) =1/(\cos \pi \gamma /4).$

Black circles:  numerical results for $p=1,2$ components;  White circles: numerical results for the $p=0$ component computed for 3 different integration contours for  $\Re e\gamma = .5,-.5,-1.5;$  White dots: ultra asymptotic saddle points at $\gamma =1/2;$ Full lines: the
functions $\chi _{p}\left( \gamma \right) $ for $\left( 1,2\right);$ Dashed lines, the function $\chi _{0}\left( \gamma \right) $ including two auxiliary branches. Arrows indicate the direction of increasing $Q.$

\bigskip
Fig.2. {\it Plot of effective intercept vs. effective dimension at
fixed $Q^{2}$}

The same as in Fig.1 but now for fixed $\ln 
Q^{2} / Q_{0}^{2}=4.$  The  results are computed  for $Y=\left\{ 4,6,8,10\right\} .$ The arrows describe increasing $Y.$

\bigskip
Fig.3. {\it Plot of effective intercept vs. effective dimension for a singular weight}

\medskip
The plot is the same as Fig.1 with a weight $f_{p}\left( \gamma \right)
\propto 1/(\gamma \cos \pi \gamma /4),$ i.e. singular at $\gamma = 0.$ Note the accumulation
of black circles near the singularity at $\mathop{\rm Re}\gamma =0$ for $p=1,2.$

\bigskip
Fig.4. {\it Comparison with  \cite{donnadrie}}

\medskip
The plot is similar to Fig.1, except for  a rescaling of the vertical coordinate $Y\rightarrow \bar{\alpha}Y,$ with $%
\bar{\alpha}=.15.$ The curves denoted $\epsilon_{0,1}$ correspond to the same rescaling of  $\chi _{0,1}\left( \gamma \right).$   The black circles correspond to our calculations at fixed
$Y=10$ and  $\ln Q^{2}/Q_{0}^{2}=\left(
4,6,8,10\right).$  The  results  for paper \cite{donnadrie} corresponding to the same values of $Y$ and  $\ln Q^{2}/Q_{0}^{2}$  are given by
crosses. The arrows indicate the direction of increasing 
 $ Q^{2}.$

\begin{center}
\psfig{figure=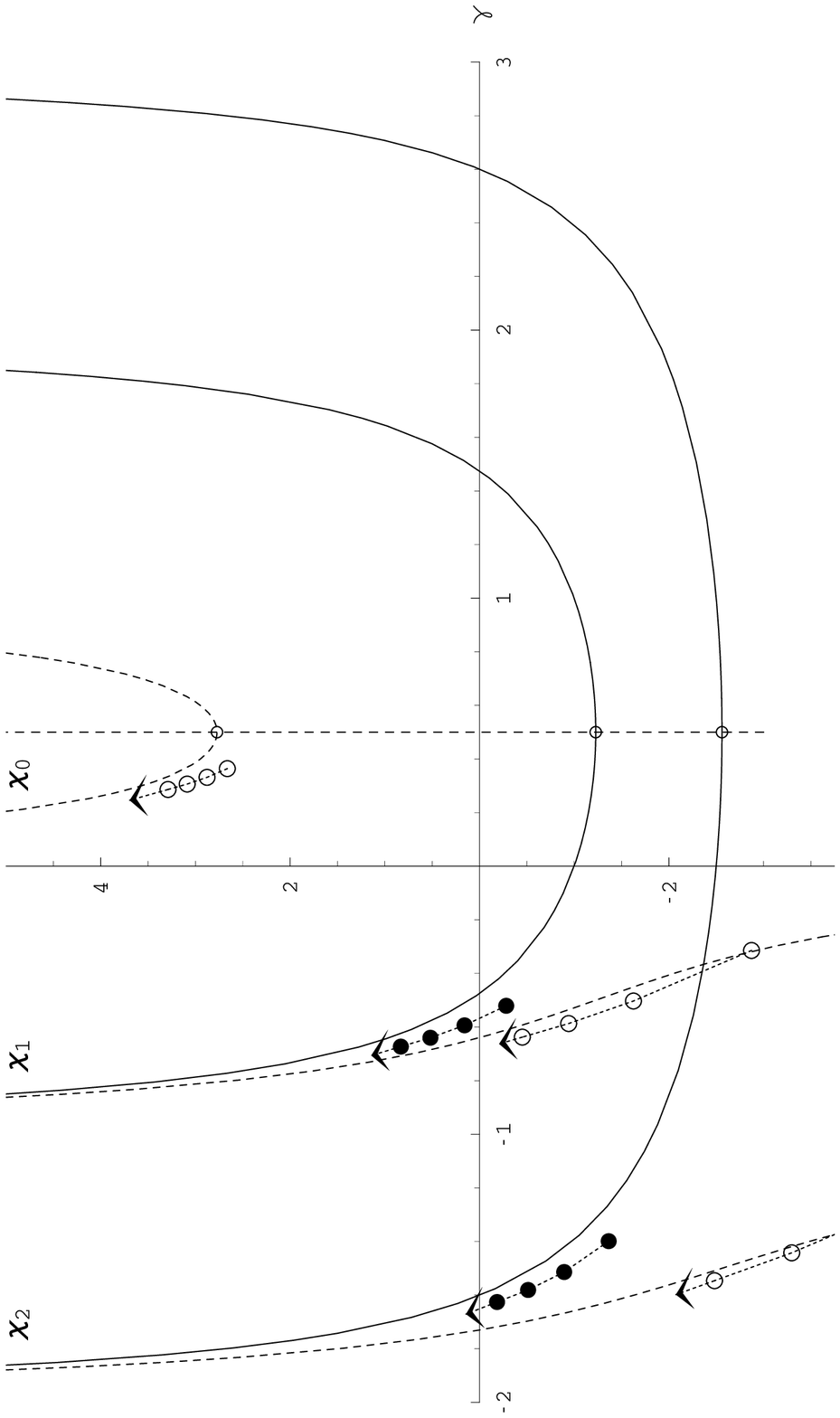,height=17cm}\\
\vskip 1cm
{\large\bf Figure 1}\end{center}

\clearpage
\begin{center}
\psfig{figure=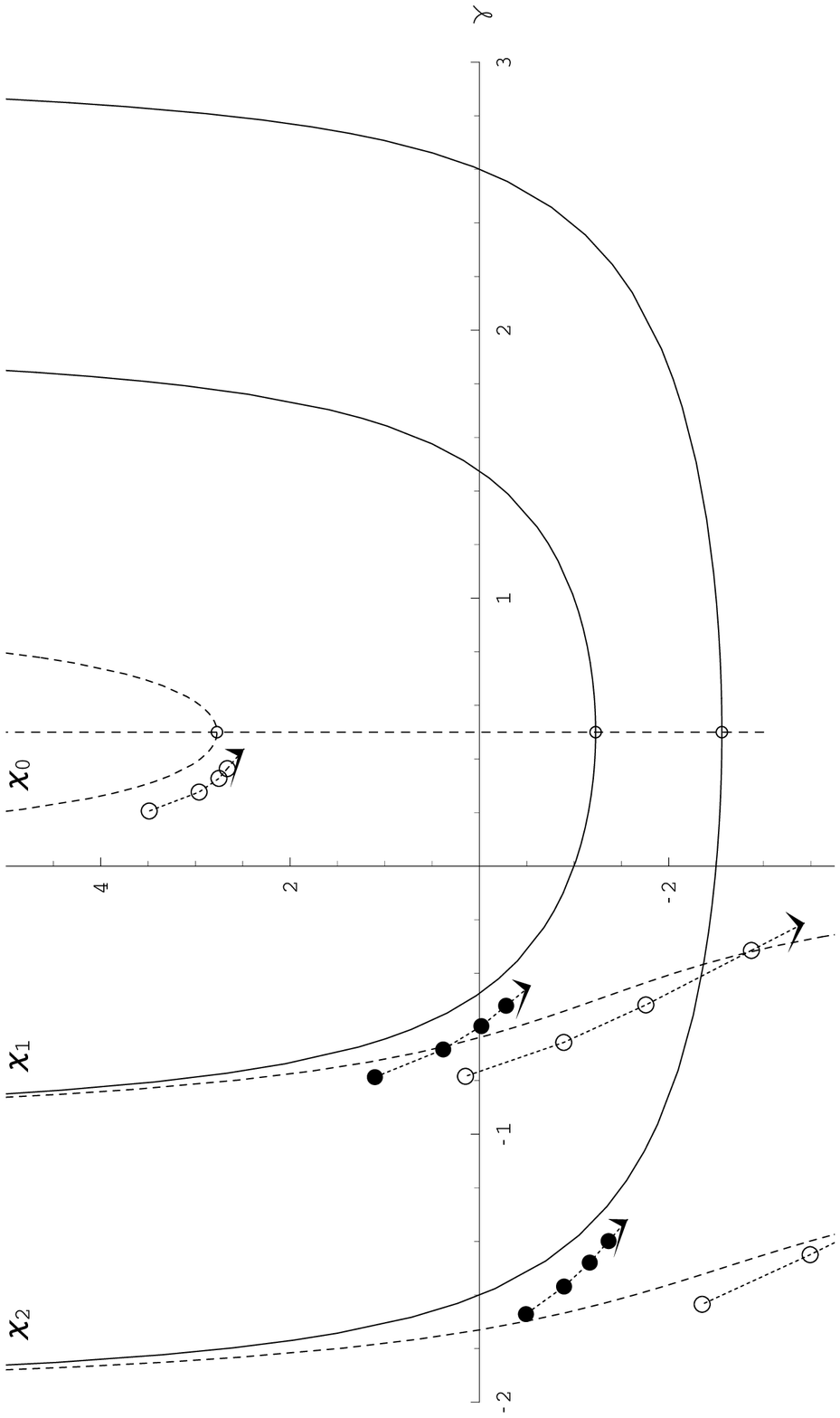,height=17cm}\\
\vskip 1cm
{\large\bf Figure 2}\end{center}

\clearpage
\begin{center}
\psfig{figure=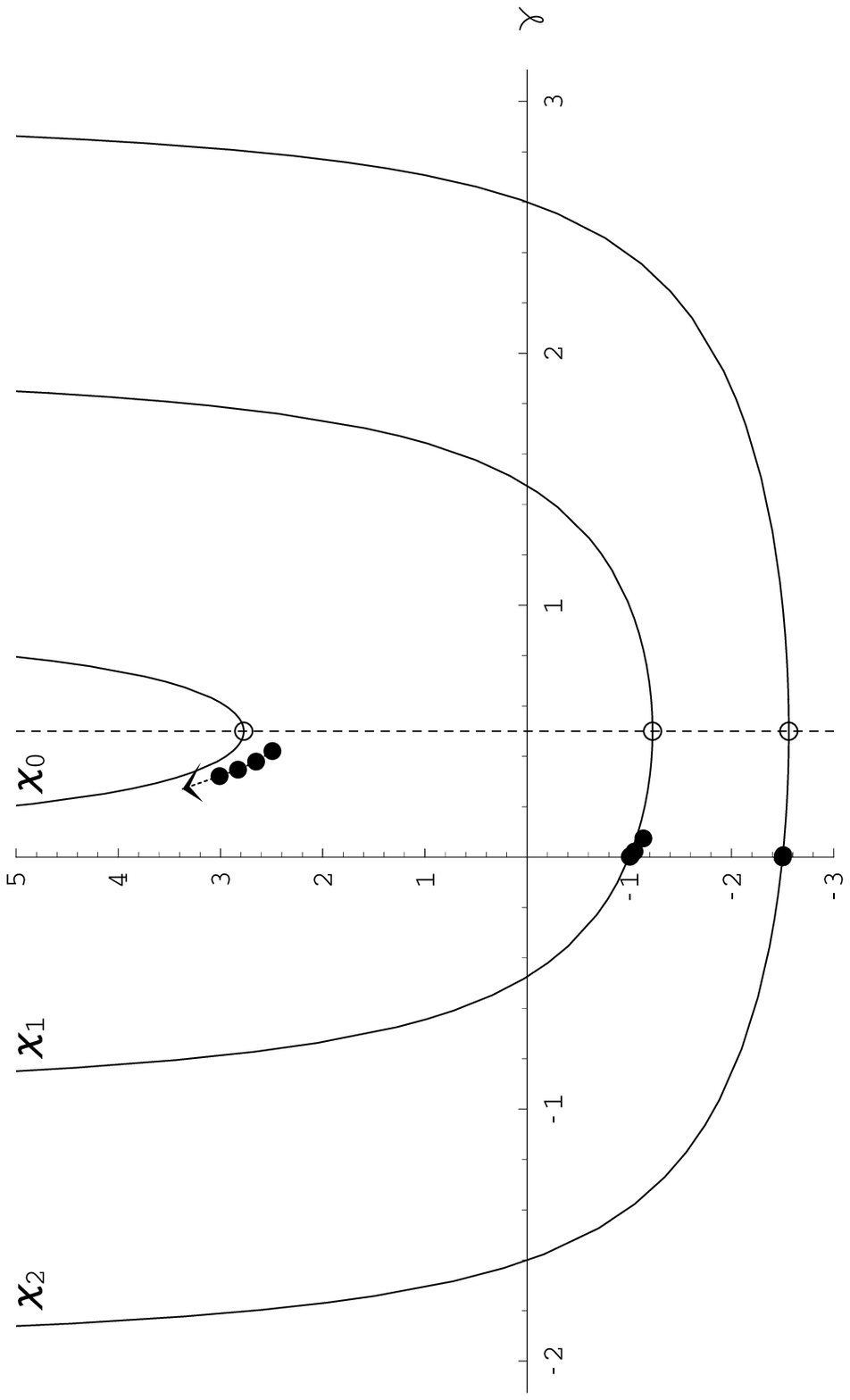,height=17cm}\\
\vskip 1cm
{\large\bf Figure 3}\end{center}

\clearpage
\begin{center}
\psfig{figure=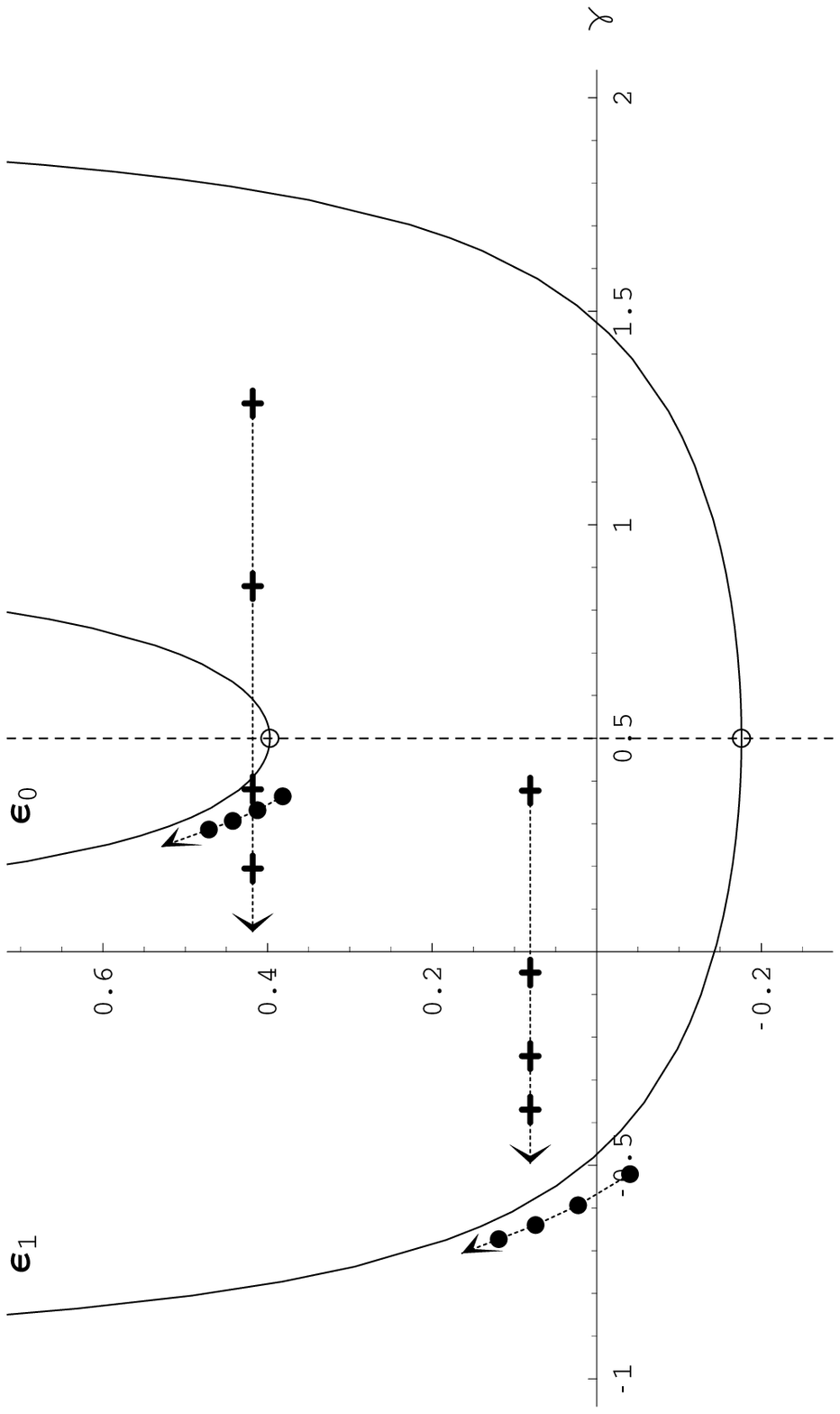,height=17cm}\\
\vskip 1cm
{\large\bf Figure 4}\end{center}


\begin{thebibliography} {99}
\bibitem{parametrization}
The parametization of Ref.[1] is the following 
\begin{eqnarray}
f_{0}\left(
Q^{2}\right) &=& A_{0}\left( \frac{Q^{2}}{Q^{2}+a_{0}}\right) ^{1+\epsilon_0}\left(
1+X\log \left( 1+\frac{Q^{2}}{Q_{0}^{2}}\right) \right) \nonumber\\
f_{1}\left( Q^{2}\right) &=& A_{1}\left( \frac{Q^{2}}{Q^{2}+a_{1}}\right)
^{1+\epsilon_{1}}\frac{1}{1+\sqrt{Q^{2}/Q_{1}^{2}}},\nonumber\\
f_{2}\left( Q^{2}\right) &=&A_{2}\left( \frac{Q^{2}}{Q^{2}+a_{2}}\right)
^{1+\epsilon_{2}},
\nonumber
\end{eqnarray}
$\begin{array}{lllll}
\epsilon_{0}=0.418 &  & \epsilon_{1}=.0808 &  & \epsilon_{2}=-0.4525 \\ 
A_{0}=0.0410 &  & A_{1}=.387 &  & A_{2}=.0504 \\ 
a_{0}=7.13 &  & a_{1}=0.684 &  & a_{2}=.00291 \\ 
Q_{0}^{2}=10.6 &  & Q_{1}^{2}=48.0 &  &  \\ 
X=0.485 &  &  &  & 
\end{array}$
\end{thebibliography}
\end{document}